\documentclass[reprint,amsmath,amssymb,aps,prx,longbibliography,%
superscriptaddress]{revtex4-1}

\usepackage[utf8]{inputenc}
\usepackage{graphicx}
\usepackage[usenames]{color}
  \definecolor{goodgreen}{rgb}{0.1,0.5,0}
  \definecolor{goodred}{rgb}{0.7,0,0}
\usepackage[colorlinks,urlcolor=blue,citecolor=goodgreen,linkcolor=goodred]%
{hyperref}

\usepackage{mathptmx}
\usepackage[scaled=.90]{helvet}
\usepackage{courier}

\newcommand{\un}[1]{\ensuremath{\,\text{#1}}}
\newcommand{\vsd}{\ensuremath{V_{\text{sd}}}}
\newcommand{\vbg}{\ensuremath{V_{\text{g}}}} 
\newcommand{\vg}{\vbg}
\newcommand{\nel}{\ensuremath{N_\text{el}}}
\newcommand{\didv}{\ensuremath{\text{d}I/\text{d}\vsd}}

\newcommand{\xzpf}{\ensuremath{x_\text{zpf}}}
\newcommand{\bpar}{\ensuremath{B_\parallel}}

\newcommand{\braket}[2]{\left.\left<{#1}\right|{#2}\right>}
\newcommand{\DSO}{{\ensuremath{\Delta_\text{SO}}}}
\newcommand{\DKK}{{\ensuremath{\Delta_\text{KK'}}}}

\begin{document}

\title{Magnetic field control of the Franck-Condon coupling of few-electron
quantum states}

\author{P. L. Stiller}
\author{A. Dirnaichner}
\author{D. R. Schmid}
\affiliation{Institute for Experimental and Applied Physics, 
University of Regensburg, 93040 Regensburg, Germany}
\author{A. K. H\"uttel}
\email[]{andreas.huettel@ur.de}
\affiliation{Institute for Experimental and Applied Physics, 
University of Regensburg, 93040 Regensburg, Germany}
\affiliation{Low Temperature Laboratory, Department of Applied Physics, 
Aalto University, Espoo, Finland}
\date{\today} 

\begin{abstract}
Suspended carbon nanotubes display at cryogenic temperatures a distinct
interaction between the quantized longitudinal vibration of the macromolecule
and its embedded quantum dot, visible via Franck-Condon conductance side bands 
in transport spectroscopy. We present data on such side bands at known absolute 
number $\nel=1$ and $\nel=2$ of conduction band electrons and consequently 
well-defined electronic ground and excited states in a clean nanotube device. 
The interaction evolves only at a finite axial magnetic field and displays a 
distinct magnetic field dependence of the Franck-Condon coupling, different for 
different electronic base states and indicating a valley-dependent 
electron-vibron coupling. A tentative cause of these effects, reshaping of the 
electronic wavefunction by the magnetic field, is discussed and demonstrated in 
a model.
\end{abstract}

\maketitle

\section{Introduction}

Vibrational degrees of freedom, typically approximated at small deflection as 
harmonic oscillators, contribute in many ways to the fundamental properties of 
matter. The Franck-Condon principle \cite{tfs-franck-1926, pr-condon-1926}
relates vibrational wave functions in molecular physics to the intensity 
envelope of vibrational side bands in optical spectra. This principle, where an 
electronic transition is assumed to be instantaneous compared to the slow 
motion of the nuclei, also becomes directly visible in electronic 
low-temperature transport spectroscopy of single (macro)molecules 
\cite{nature-park-2000, prb-braig-2003, nature-leroy-2004, prl-weig-2004, 
prl-koch-2005, njp-izumida-2005, prl-sapmaz-2006, prb-koch-2006, 
advmat-osorio-2007, nl-burzuri-2014}. An experimental system where such
Franck-Condon sidebands have been observed consistently is the longitudinal 
(stretching mode) vibration of a suspended single wall carbon nanotube quantum 
dot \cite{prl-sapmaz-2006, quantumnems, nphys-leturcq-2009, cocoset, 
nl-island-2012, nnano-ganzhorn-2013, nl-jung-2013, nl-weber-2015}. Results 
range from the nanotube length dependence of the vibration frequency 
\cite{prl-sapmaz-2006}, or thermal occupation of a vibration mode 
\cite{nphys-leturcq-2009}, all the way to electronic pumping of nonequilibrium 
occupation \cite{cocoset}, spin-vibron coupling \cite{nnano-ganzhorn-2013} and 
a spin-dependent, electrostatically tunable electron-vibron coupling 
\cite{nl-weber-2015}.

Here, we present first observations of Franck-Condon sidebands at known 
absolute number $\nel=1$ and $\nel=2$ of conduction band electrons in the
unperturbed carbon nanotube transport spectrum. The vibrational sidebands 
evolve only at finite axial magnetic field \bpar. The resulting millikelvin
transport spectrum displays different sideband behaviour depending on the 
electronic base state; the data indicates a valley-dependent Franck-Condon 
electron-vibron coupling parameter \cite{prb-braig-2003}. As a tentative 
mechanism for the observed phenomenon, reshaping of the electronic wavefunction 
by the magnetic field \cite{highfield} is discussed and modelled; the model 
manages to capture the essential observed behaviour.

\begin{figure}[t]
\centering
\includegraphics{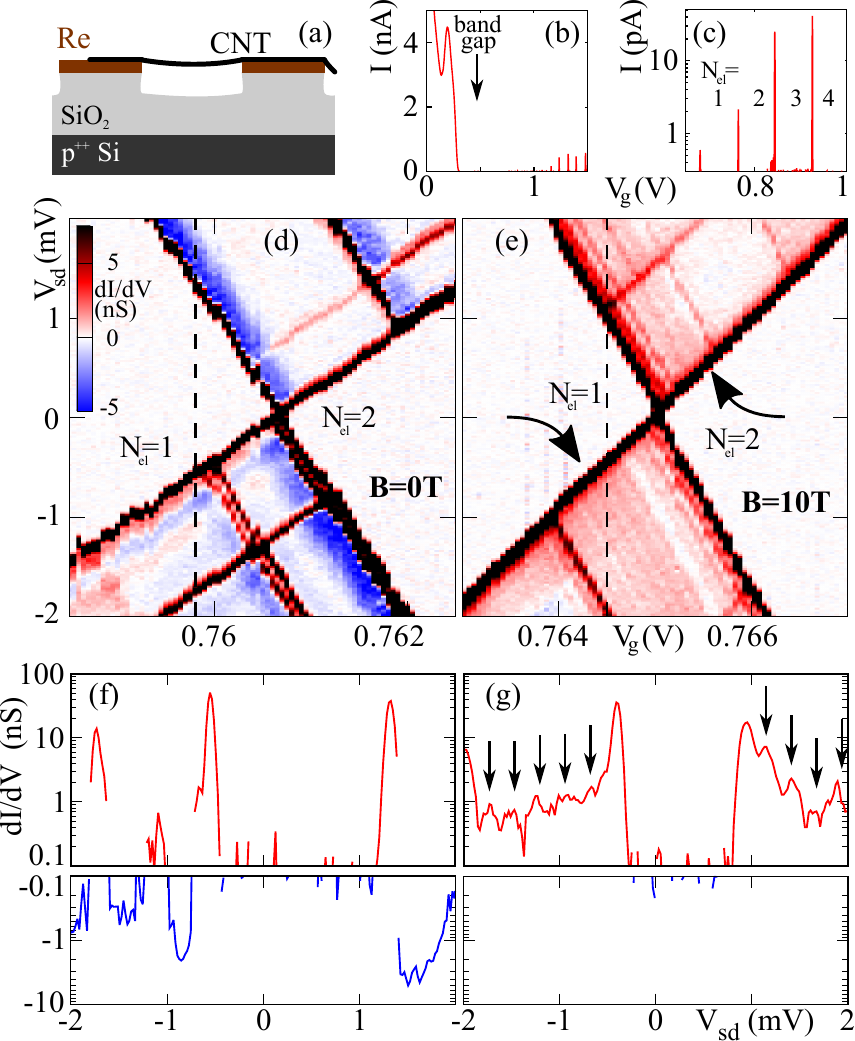}
\caption{
(a) Schematic device geometry. A carbon nanotube is grown in situ across 
pre-defined rhenium contact electrodes and a trench.
(b) Overview device characterization $I(\vg)$ at $\vsd=50\,\mu\text{V}$, 
showing transparent behaviour in hole conduction, the band gap, and Coulomb 
oscillations in electron conduction.
(c) Coulomb oscillations $I(\vg)$ for $\vsd=0.5\un{mV}$ near the band gap, with 
absolute electron numbers \nel\ marked.
(d,e) Differential conductance \didv\ at the $1\le \nel \le 2$ transition, for
a magnetic field of (d) $\bpar=0$ and (e) $\bpar=10\un{T}$ parallel to the 
nanotube axis (identical color scale, cut off at $+7\un{nS}$ for better
contrast).
(f) Trace $\didv(\vsd)$ at $\bpar=0\un{T}$, $\vg=0.7598\un{V}$, 
see dashed line in (d), in logarithmic scale. The upper panel plots regions of 
positive \didv, the lower panel regions of negative \didv.
(g) Trace $\didv(\vsd)$ at $\bpar=10\un{T}$, $\vg=0.7645\un{V}$, 
see dashed line in (e), using the same plotting method and scale as in (f). The 
equidistant arrows indicate harmonic excitation lines. 
\label{fig:overview}}
\end{figure}

\section{Conduction side bands at finite field}

A sketch of the measured device is depicted in Fig.~\ref{fig:overview}(a). 
Following \cite{nmat-cao-2005}, a carbon nanotube is grown over pre-defined 
rhenium contact electrodes and etched trenches. Subsequently the device is 
cooled down in a top-loading dilution refrigerator and characterized 
electronically at a base temperature of $T_\text{mc}\le 30 \un{mK}$, immersed 
into the diluted phase of the $^3\text{He}/{}^4\text{He}$ mixture. The length 
of the suspended nanotube segment is $L=700\un{nm}$. The device has already 
been characterized electronically in \cite{highfield, kondocharge, fewholes};
as can also be seen in the current trace at low bias of
Fig.~\ref{fig:overview}(b), it displays the behaviour of a small bandgap
single-wall carbon nanotube, with transparent hole conduction and strong
Coulomb blockade at low electron numbers. The first Coulomb oscillations 
exhibit very low current and require particular care to be resolved, see 
Fig.~\ref{fig:overview}(c). Here, the opaque tunnel barriers are given by p-n 
junctions extended along the nanotube, between the electrostatically induced 
n-quantum dot and p-behaviour near the leads \cite{apl-park-2001, 
nnano-steele-2009}.

In the following we focus on the $1\le \nel \le 2$ transition, i.e., the second
Coulomb oscillation at the electron side of the band gap. 
Fig.~\ref{fig:overview}(d) and Fig.~\ref{fig:overview}(e) display the stability
diagram close to the corresponding degeneracy point, for (d) $\bpar=0$ and for 
(e) a magnetic field $\bpar=10\un{T}$ applied in parallel to the carbon 
nanotube axis. The overall conductance is very low; even so, the color scale in 
the figure has been cut off such as to focus on the substructure of the single 
electron tunneling (SET) regions. The strong black lines, corresponding to 
electronic excitations, shift with magnetic field; their detailed behaviour, as 
well as the negative differential conductance at $\bpar=0$, is topic of ongoing 
analysis. The figures display a clear qualitative difference: while the areas 
between the electronic excitation lines are featureless in 
Fig.~\ref{fig:overview}(d), in Fig.~\ref{fig:overview}(e) they display a 
multitude of fine, closely spaced conductance resonances (see the arrows in 
Fig.~\ref{fig:overview}(e)). This also becomes visible in the comparison of the 
trace cuts of Fig.~\ref{fig:overview}(f) ($\bpar=0\un{T}$) and 
Fig.~\ref{fig:overview}(g) ($\bpar=10\un{T}$).

\begin{figure}[t]
\centering
\includegraphics{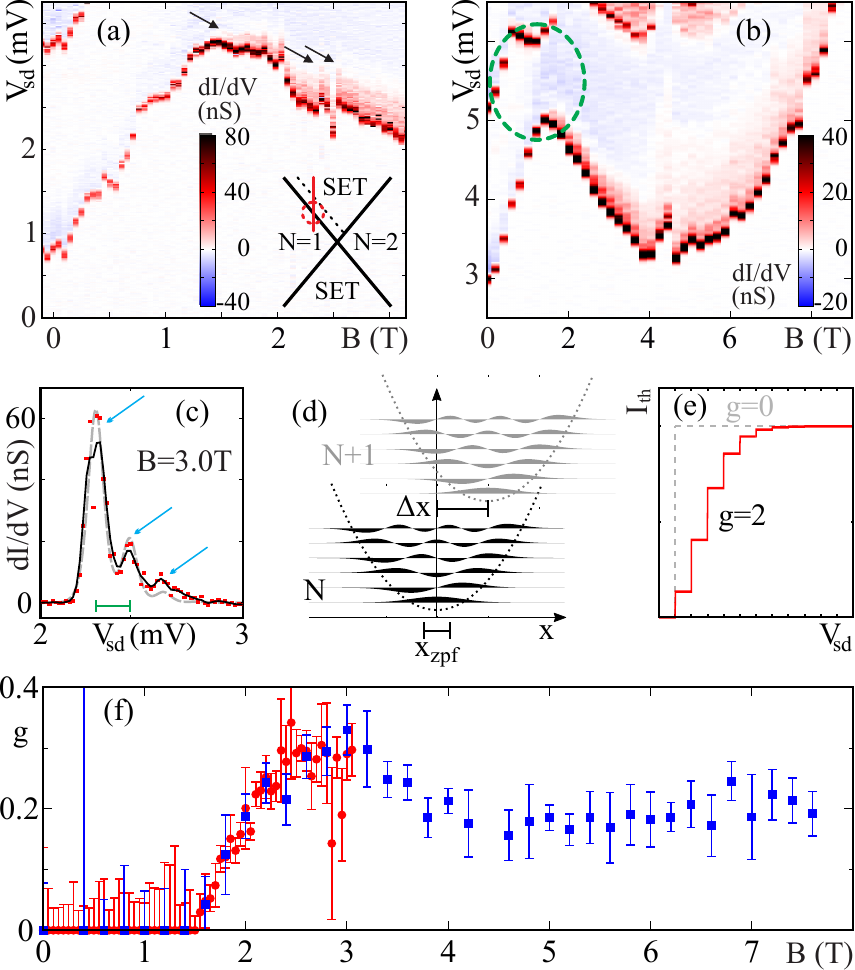}
\caption{
(a) Differential conductance as function of magnetic field parallel to the 
nanotube axis \bpar\ and bias voltage \vsd, $\didv(\bpar,\vsd)$, at constant
$\vg=0.7599\un{V}$. This cuts through the $1\le \nel \le 2$ single electron
tunneling region, as sketched in the inset. 
(b) Similar measurement of $\didv(\bpar,\vsd)$, covering a larger magnetic 
field range at lower resolution; $\vg = 0.758\un{V}$. 
(c) Solid line: smoothened trace $\didv(\vsd)$ from (a) at $\bpar=3\un{T}$, 
displaying the edge of the Coulom blockade region, corresponding to the 
$1\le \nel \le 2$ ground state transition, and conductance side peaks;
red points: unfiltered numerical derivative; gray dashed line: fit curve, see 
the text.
(d) Illustration of the Franck-Condon coupling mechanism, see the text.
(e) Schematic of the stepwise current increase at increasing bias voltage due 
to Franck-Condon coupling, here for $\Delta x = 2 x_\text{zpf}$ and thus $g=2$ 
as in (d).
(f) Franck-Condon coupling parameter $g(\bpar)$ as function of the magnetic 
field; see Appendix~\ref{a-fit} for details of the fit procedure. Red dots: 
data of (a); blue squares: data of (b).
\label{fig:groundstate}} 
\end{figure}
Figure~\ref{fig:groundstate}(a) demonstrates the emergence of this
phenomenon with increasing magnetic field. Here, we show the differential 
conductance $\didv(\bpar,\vsd)$ as function of an applied field \bpar\ parallel 
to the carbon nanotube axis and of the bias voltage \vsd. The gate voltage is 
kept constant and chosen such that we trace across the $\nel = 1$ edge of the 
$1\le \nel \le 2$ single electron tunneling (SET) region, see the inset of 
Fig.~\ref{fig:groundstate}(a) for a sketch and Fig.~\ref{fig:harmonic}(a) in 
Appendix \ref{app-harm} for a larger-scale plot of the conductance at the 
degeneracy point. Here and later, we focus our evaluation on the low 
magnetic field region since it provides a better signal/noise ratio.

For $\bpar=0$, the SET region edge, visible as line of differential 
conductance, is located at approximately $\vsd = 0.8\un{mV}$. Due to a shift in 
energy of the electronic states involved in transport, it rapidly moves to 
higher bias voltages until $\bpar \simeq 1.5\un{T}$ is reached. Here, the 
magnetic field induces a change in ground state, leading to a different energy 
dispersion. Soon afterwards, side bands of the differential conductance line 
emerge, see the arrows in Fig.~\ref{fig:groundstate}(a). 
Figure~\ref{fig:groundstate}(b) displays a larger parameter range than 
Fig.~\ref{fig:groundstate}(a), though measured at reduced resolution. Still, 
the side bands of the conductance line become clearly visible as an asymmetric 
broadening of the main conductance line (towards higher bias voltages).

An example trace cut from Fig.~\ref{fig:groundstate}(a), both smoothened for 
clarity (black line) and as raw numerical derivative of the current (red 
points), is shown in Fig.~\ref{fig:groundstate}(c). A manual analysis of the 
relative peak positions in each such recorded trace $I(\vsd)$ is given in 
Appendix~\ref{app-harm}, see in particular Fig.~\ref{fig:harmonic}(f). Its 
conclusion is that within the scatter the side bands are equidistant within 
each trace. The excitation energy is magnetic field independent for $\bpar 
\lesssim 6\un{T}$ \footnote{At higher fields $\bpar \gtrsim 6\un{T}$ the 
extracted peak positions may indicate an increase in oscillator quantum, though 
the signal/noise ratio makes a quantitative analysis challenging here.} and 
given by $\Delta \epsilon \simeq 50 \, \mu\text{eV}$. This indicates a harmonic 
oscillator independent of the electronic spectrum. Given its energy scale, we 
can tentatively identify it with the longitudinal vibration of the carbon 
nanotube \cite{prl-sapmaz-2006}.

\section{Franck-Condon model}

In multiple publications, the mechanism leading to vibrational harmonic side 
bands in transport spectroscopy has been identified as the Franck-Condon 
principle \cite{prb-braig-2003, prl-koch-2005, prl-sapmaz-2006}. As sketched in
Fig.~\ref{fig:groundstate}(d), the equilibrium position of the vibrational
harmonic oscillator depends on the number of charges \nel\ on the nanotube; the
rate of single electron tunneling through its quantum dot is modified by the
spatial overlap of the involved harmonic oscillator states $\left|
\braket{\Psi_m(N)\,}{\Psi_n(N+1)} \right|^2 = \left|
\braket{\Psi_m(x)\,}{\Psi_n(x+\Delta x)} \right|^2$, with $m$ and $n$ as
the vibrational quantum number at $N$ and $N+1$ electrons, respectively. 
$\Delta x$ is the displacement of the harmonic oscillator by the
additional charge, cf. Fig.~\ref{fig:groundstate}(d). The coupling strength is
parametrized via the Franck-Condon coupling parameter $g = (\Delta x / \xzpf)^2 
/ 2$, comparing $\Delta x$ with the characteristic length scale of the harmonic 
oscillator $\xzpf = \sqrt{\hbar / m \omega}$.

As sketched in Fig.~\ref{fig:groundstate}(e), a finite value of $g$ leads to a 
redistribution of current between vibrational state channels: at the bias 
voltage corresponding to the bare electronic transition energy $N 
\longrightarrow N+1$ current is suppressed, but it increases whenever an 
additional vibration state becomes energetically available. A large number of 
extensions to this model has been developed to take into account specific 
details of transport spectra, see, e.g.,  \cite{njp-izumida-2005, 
prb-zazunov-2006, prl-haertle-2009, prb-vonoppen-2009, prb-mariani-2009, 
prb-huebener-2009, prb-cavaliere-2010, prb-maier-2011, prb-haertle-2011, 
prb-yar-2011, prb-fang-2011, njp-donarini-2012, prb-zhang-2012, 
prb-jovchev-2013, jchemp-simine-2014, nl-weber-2015, prb-mccaskey-2015, 
prb-mol-2017}; however, for now we focus our analysis on the 
simplest theoretical case, assuming a single harmonic oscillator mode and fast 
relaxation into the vibrational ground state. In this case, the current step 
heights or conductance peak amplitudes follow the Poisson formula 
\cite{prb-braig-2003, prb-koch-2006}, $P_n = {(e^{-g} g^n)}/{n!}, \quad n=0, 1, 
2, \dots$, at an energy $\Delta E_\text{vib}= n \hbar \omega$ from the bare 
electronic state transition supplied via the bias voltage. The resulting step 
function of the current, in absence of broadening effects, is sketched in 
Fig.~\ref{fig:groundstate}(e) for the example of a large $g=2$. 

At base temperature, with $25\un{mK} \, k_\text{B} \simeq 2\,\mu\text{eV}$, we 
expect the conductance lines to be lifetime-broadened, with a Lorentzian line 
shape. Thus, a sum of Lorentzians with amplitudes following the Poisson 
sequence and center points shifted by equidistant bias voltage offsets can be
envisioned as model. In practice, it turns out that the Lorentzian line shape 
does not suit the measurement data well; this may indicate broadening effects 
beyond temperature and lifetime, as, e.g., fluctuating gate or bias voltages. 
Empirically, we choose the resonance shape of thermal broadening, $\propto 
\cosh^{-2}$, instead \cite{prb-beenakker-1991}. Details of the fit procedure can 
be found in Appendix~\ref{a-fit}.

The result of evaluating the Franck-Condon coupling parameter $g$ for each 
trace $\didv(\vsd)$ at fixed \bpar\ is plotted in 
Fig.~\ref{fig:groundstate}(f). It shows the resulting magnetic field dependence 
$g(\bpar)$ for the $\nel=1$ $\longrightarrow$ $\nel=2$ ground state transition. 
The absence of side bands for $\bpar < 1.5\un{T}$ corresponds to an absence of 
coupling, i.e., $g \sim 0$. For $\bpar \ge 1.5\un{T}$ the coupling increases 
monotonously, reaching a maximum value $g(\bpar) \simeq 0.3$ at $2.5T \le \bpar 
\le 3\un{T}$. Subsequently, we observe a slow decrease and stabilization 
at $g(\bpar) \simeq 0.2$.

\section{Relation to electronic states}

To our best knowledge, no similar observations of a magnetic field dependent 
electron-vibron coupling have been published so far. Its onset at an
anticrossing of electronic states, see the dashed ellipsoid in
Fig.~\ref{fig:groundstate}(b), suggests a connection to the electronic quantum 
numbers. As opposed to previous reports on the longitudinal vibration mode 
\cite{prl-sapmaz-2006, quantumnems, nphys-leturcq-2009, cocoset,
nnano-ganzhorn-2013, nl-weber-2015}, here we are characterizing a device where
the nanotube has grown cleanly across pre-existing contacts, and no (accidental
or intentional) strongly inhomogeneuous potential distorts the wave functions 
in the suspended macromolecule away from the metallic contacts 
\cite{highfield}. With this in mind, we have analyzed the vibrational side band 
behaviour of the electronic {\it excited} states in the transport spectrum.

\begin{figure}[t]
\centering
\includegraphics{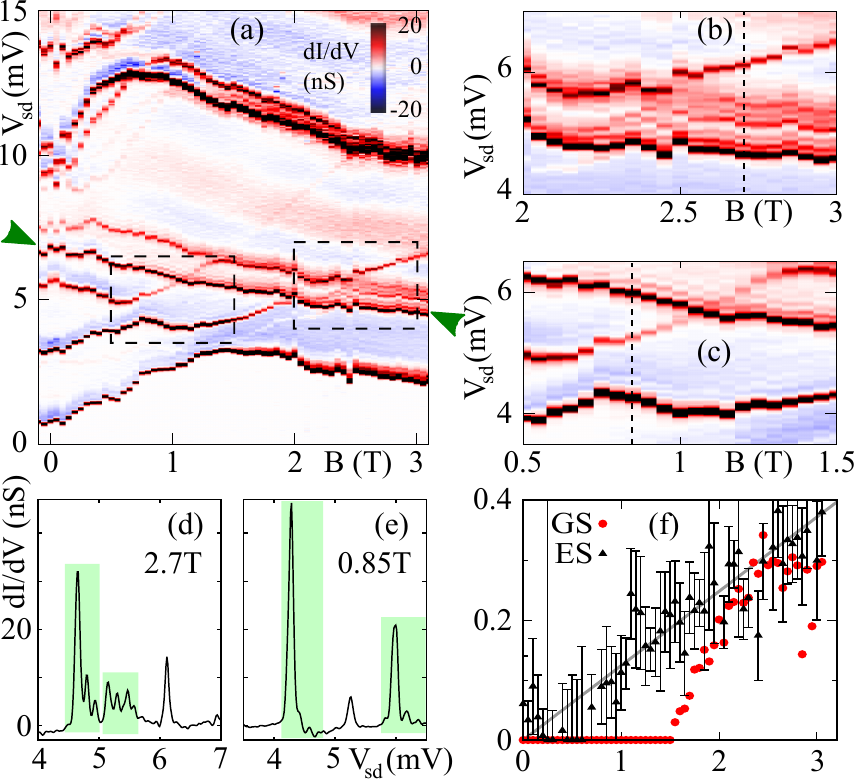}
\caption{
(a) $1\le \nel \le 2$ excitation spectrum: Differential conductance 
$\didv(\bpar, \vsd)$ as function of magnetic field \bpar\ and bias voltage 
\vsd, for constant gate voltage $\vg=0.7599\un{V}$.
(b), (c) Detail enlargements (same color scale) of the areas marked with dashed
rectangles in (a).
(d), (e) Trace cuts $\didv(\vsd)$ at the positions indicated in (b) and
(c). Conductance peaks with clear vibrational side bands are shaded.
(f) Magnetic field evolution of the Franck-Condon coupling $g(\bpar)$ for
the excited state transition marked with arrowheads in (a) (black), compared 
with the ground state transition (red dots, same data as in
Fig.~\ref{fig:groundstate}(f)). For the ground state the error bars (already 
shown in Fig.~\ref{fig:groundstate}(f)) have been omitted for clarity. The 
solid line is a linear fit after manual removal of outliers.
\label{fig:excitations}} 
\end{figure}
A plot of the differential conductance at fixed gate voltage, as function of
\bpar\ and the bias voltage \vsd, now over a large bias range, is shown in
Fig.~\ref{fig:excitations}(a). A preliminary evaluation of the data of
Fig.~\ref{fig:excitations}(a), taking also into account the one-electron
excitation spectrum of the device \cite{highfield}, reveals two energetically
close shells with both intra- and inter-shell exchange interaction 
\cite{marganska}; detailed modelling of the two electron transport spectrum 
will be the topic of a separate work. Here, we limit ourselves to a 
straightforward classification of conductance lines by magnetic field 
dispersion; see Appendix~\ref{app-classif} for the details. 

The dominant magnetic field dependence of the electron energies in a carbon 
nanotube in an axial field originates from the electronic orbital magnetic 
moment $\mu_\text{orb}$, see, e.g., \cite{nature-minot-2004, rmp-laird-2015, 
highfield}, and Appendix~\ref{app-classif}. Thus, both one- and two-electron 
quantum states become at large field angular momentum (and valley) eigenstates, 
and the slope of a conductance line in Fig.~\ref{fig:excitations} indicates the 
angular momentum {\em change} when a second electron tunnels onto the quantum 
dot. If the state of the first electron remains unchanged, only the 
contribution of the second electron to the magnetic moment --- parallel or 
antiparallel to the magnetic field --- determines the magnetic field 
dispersion. From this we can classify the conductance lines of 
Fig.~\ref{fig:excitations}(a). As expected, in the low-bias region of the 
figure, two dominant slopes clearly emerge; we identify these with the addition 
of a K'-valley electron (downward slope) or a K-valley electron (upward slope), 
respectively \cite{nature-minot-2004, rmp-laird-2015, highfield}.

Figures~\ref{fig:excitations}(b) and \ref{fig:excitations}(c) enlarge the
regions marked in \ref{fig:excitations}(a) with dashed rectangles. Also here, 
the harmonic side bands are immediately visible. However, at a first glance, 
only the down-sloping spectral lines, where an electron is added in a K'-valley 
state, seem to exhibit electron-vibron coupling. Also the trace cuts at $\bpar 
= 2.70\un{T}$ (Fig.~\ref{fig:excitations}(d)) and $\bpar = 0.85\un{T}$ 
(Fig.~\ref{fig:excitations}(e)) demonstrate this, with the resonances 
accompanied by sidebands highlighted in green. 

Extracting the Franck-Condon coupling parameter $g(\bpar)$ for an exemplary 
excited state resonance (green arrowheads in Fig.~\ref{fig:excitations}(a)), 
using an identical fit procedure as for Fig.~\ref{fig:groundstate}(f), we 
compare its evolution over the entire field range $0\un{T} \lesssim B \lesssim 
3\un{T}$ with the two-electron ground state transition in 
Fig.~\ref{fig:excitations}(f). A finite $g$ persists to much lower magnetic 
field for the excited state transition; indeed, the plot shows a linear growth 
of $g(\bpar)$ in this range with $g(\bpar) \simeq 0.124 \un{1/T}\bpar$ (gray 
solid line). The sudden onset of coupling for the ground state transition is 
thus consistent with a valley dependence and the change in ground state quantum 
numbers at $\bpar \simeq 1.5\un{T}$, see the structure marked with a dashed 
ellipsoid in Fig.~\ref{fig:groundstate}(b).

\begin{figure}[t]
\centering
\includegraphics{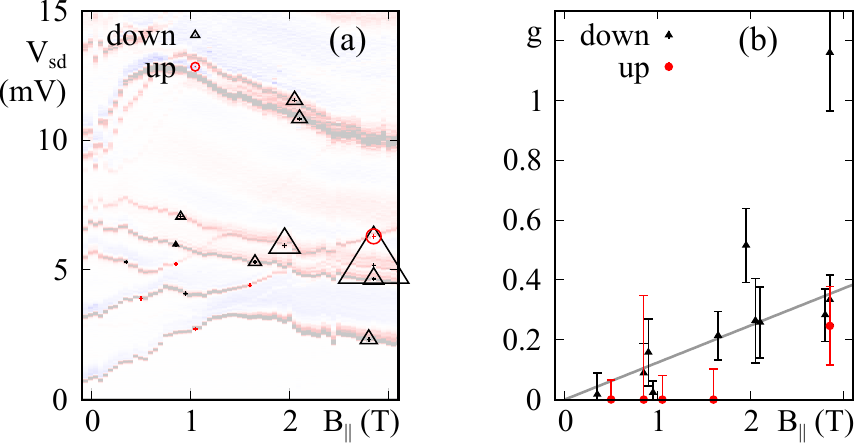}
\caption{Evaluation of the conductance resonances in 
Fig.~\ref{fig:excitations}(a) for harmonic side bands; same data plotted in two 
different representations: (a) Coupling $g$ as symbol size (downward lines, 
black triangles; upward lines, red circles) for the parameter set (\bpar,\vsd) 
where it was evaluated, and superimposed on the resonance pattern (background); 
(b) coupling $g$ as function of magnetic field \bpar, irrespective of bias 
voltage. The solid line is the linear fit of Fig.~\ref{fig:excitations}(f).
\label{fig:updown}} 
\end{figure}
We have performed a systematic analysis of the conductance resonances in 
Fig.~\ref{fig:excitations}(a), by dividing them into segments at each crossing 
or anticrossing and evaluating each segment; details can be found in 
Appendix~\ref{app-excited}. The result is shown in Fig.~\ref{fig:updown}. 
Fig.~\ref{fig:updown}(a) displays symbols with their size representing the 
obtained $g$-parameter, superimposed on the line pattern of 
Fig.~\ref{fig:excitations}(a) at the parameters $(\bpar, \vg)$ where the 
evaluation has taken place. Fig.~\ref{fig:updown}(b) plots the value of $g$ as 
function of the selected magnetic field, irrespective of bias voltage \vsd\ 
and thus selected resonance line, with the fit function of 
Fig.~\ref{fig:excitations}(f) added. In both cases, red circles represent 
downsloping lines, black triangles upsloping lines. The result confirms our 
visual impression. In general the up-sloping lines, where an electron is added 
in a K-state, display no vibrational side bands, with the notable exception of 
the segment at ($3\un{T}$, $6\un{mV}$). The down-sloping lines, where an 
electron is added in a K'-state, show a $g$ growing with magnetic field, in 
most cases following the same linear behaviour as already observed in 
Fig.~\ref{fig:excitations}(f). One resonance line, also enlarged in 
Fig.~\ref{fig:excitations}(b), deviates with a large $g$.

\begin{figure}[t]
\centering
\includegraphics[width=\columnwidth]{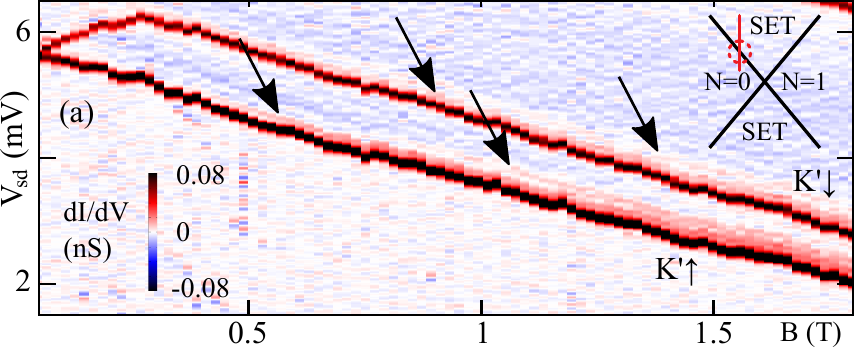}
\caption{Magnetic field behaviour of the $0 \le \nel \le 1$ one electron ground 
state transition. Traces of vibrational side bands are indicated by arrows. 
Note that the color scale is cut off at $\pm 80\un{pS}$.
\label{fig:oneelectron}} 
\end{figure}
It remains to clarify whether the observed effect is specific to the $1 \le 
\nel \le 2$ region. Figure~\ref{fig:oneelectron} displays the magnetic field 
dependence of a cut across the $0 \le \nel \le 1$ ground state transtion. Also 
here, where an electron tunnels into a K' state of the otherwise unoccupied 
conduction band, an onset of harmonic side bands can be visually identified, 
for both spin alignments. Given the significantly lower conductance here, a 
more detailed evaluation turns out to be challenging. We can however conclude 
that the electron-vibron coupling is already inherent to single electron 
phenomena, and not limited to two-electron states \cite{nl-weber-2015}. This 
indicates a selectivity directly related to the single-particle valley quantum 
number.

\section{Discussion}

The transport spectra were measured with the carbon nanotube immersed into the 
$^3\text{He} / {}^4\text{He}$ mixture (D phase) of the dilution refrigerator. 
Its viscosity at base temperature, $\eta \sim 10^{-5}\un{N s/m$^2$}$ 
\cite{book-enss}, is sufficiently high to mechanically dampen the transversal 
vibration mode \cite{heliumdamping}. As observed here, this does not affect the 
longitudinal mode; a likely explanation is that motion along the nanotube axis 
does not require any displacement of liquid.

In literature, Weber {\it et al.} \cite{nl-weber-2015} have reported an 
electronic state dependent Franck-Condon coupling in a $\nel=4n+2$ quantum dot, 
switchable via a local gate potential deformation. They apply a magnetic field 
perpendicular to the nanotube, and see no effect of that field on the 
vibrational side bands. Due to valley mixing and the field direction, their 
data is in the regime of bonding and antibonding valley linear combinations. A 
different electron-vibron coupling of (valley-ground state) spin singlet and 
(valley-distributed) spin triplet states, as observed there, obviously can 
also indicate a valley-dependent effect. Whether this is related to our results 
still requires further analysis.

Models discussing strong electron-vibron coupling in carbon nanotube quantum 
dots typically assume an inhomogeneuous charge distribution relative to the 
vibration mode envelope, or more generally, a different localization of electron 
and vibron wave function, see, e.g., \cite{prl-sapmaz-2006, prb-mariani-2009, 
prb-cavaliere-2010, njp-donarini-2012}. This is consistent with the occurrence 
of Franck-Condon side bands in devices with local gates close to the nanotube 
\cite{nphys-leturcq-2009, nl-jung-2013, nl-weber-2015}. In previous devices 
without such an electrode, a local potential may have been introduced 
via fabrication defects or impurities on the nanotube \cite{prl-sapmaz-2006, 
quantumnems, cocoset}. These expanations do not lend themselves for the device 
presented here, displaying a highly regular one electron spectrum 
\cite{highfield} and a regular addition spectrum over a large electron number 
range \cite{brokensu4, kondocharge}. Nevertheless, the formation of the quantum 
dot via gate-induced p-n junctions, see, e.g., \cite{apl-park-2001, 
nnano-steele-2009}, leads to an electronic confinement more narrow than the 
mechanically active device length, and typically centered within the suspended 
nanotube segment. 

Comparison with existing theoretical \cite{prb-mariani-2009, njp-donarini-2012} 
and experimental works \cite{nl-weber-2015} then suggests that the axial 
magnetic field modifies the Franck-Condon coupling $g$ by shifting the 
electronic wave function relative to the vibron envelope. A valley-dependent 
mechanism having this effect was recently proposed in \cite{highfield}, based 
on \cite{prb-akhmerov-2008, rmp-castroneto-2009, prb-marganska-2011}. 
Essentially, the axial magnetic field introduces an Aharonov-Bohm phase around 
the nanotube. Due to the bipartite graphene lattice, the axial component
$\kappa_\parallel$ and the circumferential component $\kappa_\perp$ of an 
electron state wave vector are coupled; one example solution is plotted in 
Fig.~\ref{fig:toymodel}(a) following \cite{highfield}. The axial 
magnetic field thus also modifies the longitudinal boundary conditions and 
thereby the charge distribution along the carbon nanotube axis: the wave 
function envelope for each graphene sublattice is tuned from a $\lambda/4$ 
shape, with a finite value at one edge, near $\bpar=0$ to a $\lambda/2$ shape, 
as expected for a traditional ``quantum box'', at large field; see the 
drawings in Fig.~\ref{fig:toymodel}(a). 

\section{Variable coupling model}
\begin{figure}[t]
\centering
\includegraphics{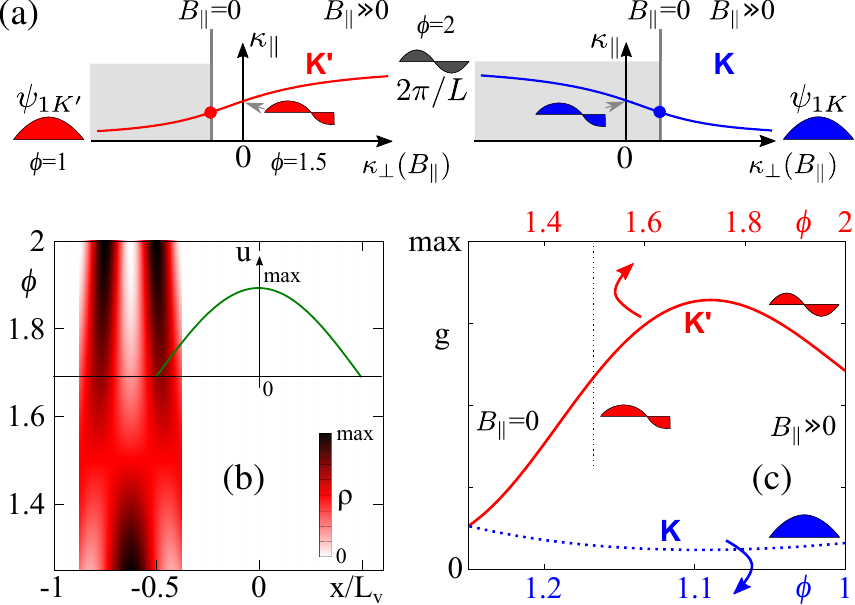}
\caption{Variable coupling model construction. (a) Low-energy wave function 
solutions for electrons on one of the carbon nanotube sublattices, in the case 
of hard longitudinal confinement and cross-quantization, following 
\cite{highfield}. Solid lines plot the allowed wave vector $(\kappa_\perp, 
\kappa_\parallel)$, schemata the resulting envelope of the sublattice wave 
function for a K' and a K state. An axial magnetic field modifies the boundary 
conditions, tuning $\kappa_\perp$ towards higher values. 
(b) Example arrangement of a vibron and a quantum dot. Green line: Vibron 
amplitude envelope, length $L_v=1$. Color plot, charge density of a quantum dot 
of length $L_d= L_v/2$, summed over both sublattices, as function of position 
and a parameter $\phi$ that tunes the per-sublattice wave function shape from 
one antinode ($\phi=1$) to two antinodes ($\phi=2$). 
(c) Approximation for the Franck-Condon parameter $g$ as function of $\phi$ 
and thereby the magnetic field \bpar, see the text.
\label{fig:toymodel}} 
\end{figure}

Based on the arguments brought forward in the previous section, we have 
constructed a toy model following \cite{highfield, njp-donarini-2012}. Details 
can be found in Appendix \ref{app-toy}; our steps are summarized as follows: we 
simplify the electronic wave function amplitudes on the two graphene 
sublattices to be of the shape $\psi\propto \sin(\pi \nu x)$ over the length of 
the quantum dot, where $\phi \in [1, 2]$ is used to continuously tune the wave 
function shape from one antinode to two antinodes and thus approximate the 
impact of the magnetic field. The total linear charge density $\rho(x)$ is 
obtained by summing up the charge densities $\propto |\psi|^2$ of the 
sublattices, one being the spatial mirror image of the other \cite{highfield}. 
The vibron is equally simplified as having a deflection amplitude $u(x) \propto 
\sin(\pi x)$.

Given the strong confinement of our electronic states and the low 
characteristic energy of our harmonic excitations, the quantum dot occupies 
likely a smaller part of the nanotube than the vibron. 
Fig.~\ref{fig:toymodel}(b) shows both charge density and vibron envelope for a 
length ration $L_d / L_v = 1/2$ and for a shift in center position $x_d - 
x_v=-0.625 \, L_v$. The charge density is plotted as function of both the 
position and $\phi$, for the range $\phi \in [1.25, 2]$ approximately covered 
by a K' state when the axial magnetic field is tuned from zero to large values, 
cf. Fig.~\ref{fig:toymodel}(a) and \cite{highfield}.

Following \cite{njp-donarini-2012}, we assume the interaction energy between 
linear charge density and deformation potential to be
\begin{equation}
 E_\text{ev} \propto \int \rho(x)\frac{\text{d}u}{\text{d}x}\,\text{d}x
\end{equation}
and the Franck-Condon parameter to be $g \propto |E_\text{ev}|^2$. 
The result for $g(\phi)$ is plotted in Fig.~\ref{fig:toymodel}(c), see the red 
solid line. Recalling that $\phi=2$ here describes the asymptotic limit of 
large \bpar, the result qualitatively agrees with the behaviour of $g$ for a K' 
state in our experiment, cf. Fig.~\ref{fig:excitations}(f) for the low-field 
and Fig~\ref{fig:groundstate}(f) for the high-field behavior. For the 
same magnetic field range, a K state covers the range $\phi \in [1, 1.25]$, 
with the large field limit $\phi=1$. Here, the coupling further decreases and 
remains small, see the dotted blue line.

Despite the many simplifications, this model describes the essential 
observations of the measurement. For a K' state, the Franck-Condon coupling 
increases from near-zero with magnetic field, reaches a maximum, and then for 
large field becomes constant at smaller value. For a K state, the coupling 
remains low. The sudden onset of coupling for the ground state around 
$\bpar=2\un{T}$, Fig.~\ref{fig:groundstate}(f), is due to a transition between 
these two cases. Limitations of the model become clear, however, when we look 
at the size ratio and placement of electron and vibron. In an ultraclean, 
suspended nanotube device, we expect the electronic system to be approximately 
centered in the suspended part. The vibron should cover at least the entire 
suspended part, which speaks against a spatial arrangement as shown in 
Fig.~\ref{fig:toymodel}(b). A parameter search using our toy model has not 
yielded fundamentally different situations with suitable $g(\bpar)$ behaviour 
so far. This shall likely require a more realistic and in particular 
also quantum mechanical treatment.

\section{Conclusions}

In conclusion, we demonstrate that vibrational sidebands emerge in the $0\le 
\nel \le 2$ transport spectrum of an ultraclean nanotube at a finite magnetic
field parallel to the nanotube axis. The sidebands are equidistant, with an 
oscillator quantum field-independent for $\bpar \lesssim 3\un{T}$. Their
evaluation results in a field-dependent Franck-Condon coupling parameter
$g(\bpar)$. Our data indicate that predominantly conductance lines 
corresponding to the addition of a K' valley electron develop a finite 
coupling parameter $g$. A tentative mechanism for this can be a field-induced 
and valley-selective modification of the electronic wavefunction envelope 
\cite{highfield}. Following \cite{njp-donarini-2012}, we have developed a 
simplified classical model and are able to reproduce the essential behaviour of 
$g$. A realistic match of device and model parameters will require a more 
detailed theoretical treatment.

It has long been proposed that, similar to existing experiments using the 
transversal vibration \cite{nphys-benyamini-2014}, also the longitudinal 
vibration coupling in a carbon nanotube Franck-Condon system can be controlled 
via changing the charge distribution along the nanotube \cite{prb-mariani-2009, 
njp-donarini-2012}. With our experimental results supporting this idea, the 
integration of the longitudinal vibration mode into quantum technological 
applications and its targeted manipulation as quantum harmonic oscillator 
\cite{cocoset} now becomes an interesting challenge.

\begin{acknowledgments}
We would like to thank M.\ Marga\'nska, M.\ Grifoni, A.~Donarini, Ch.\ Strunk,
E.\ A.\ Laird, and P.\ Hakonen for insightful discussions, and Ch.\ Strunk and 
D.\ Weiss for the use of experimental facilities. The data has been recorded 
using the Lab::Measurement software package \cite{labmeasurement}. The authors 
acknowledge financial support by the Deutsche Forschungsgemeinschaft (Emmy 
Noether grant Hu 1808/1, GRK 1570, SFB 689, SFB 1277) and by the 
Studienstiftung des deutschen Volkes.
\end{acknowledgments}

\section*{Author contributions}

The device was fabricated by P.~L.~S. and D.~R.~S; the low-temperature 
measurements were performed by all authors jointly. The resulting data was 
evaluated by P.~L.~S. and A.~K.~H.; the manuscript was written and the project 
was supervised by A.~K.~H.

\appendix

\section{Harmonicity and excitation energy}\label{app-harm}
\begin{figure*}[t]
\centering
\includegraphics{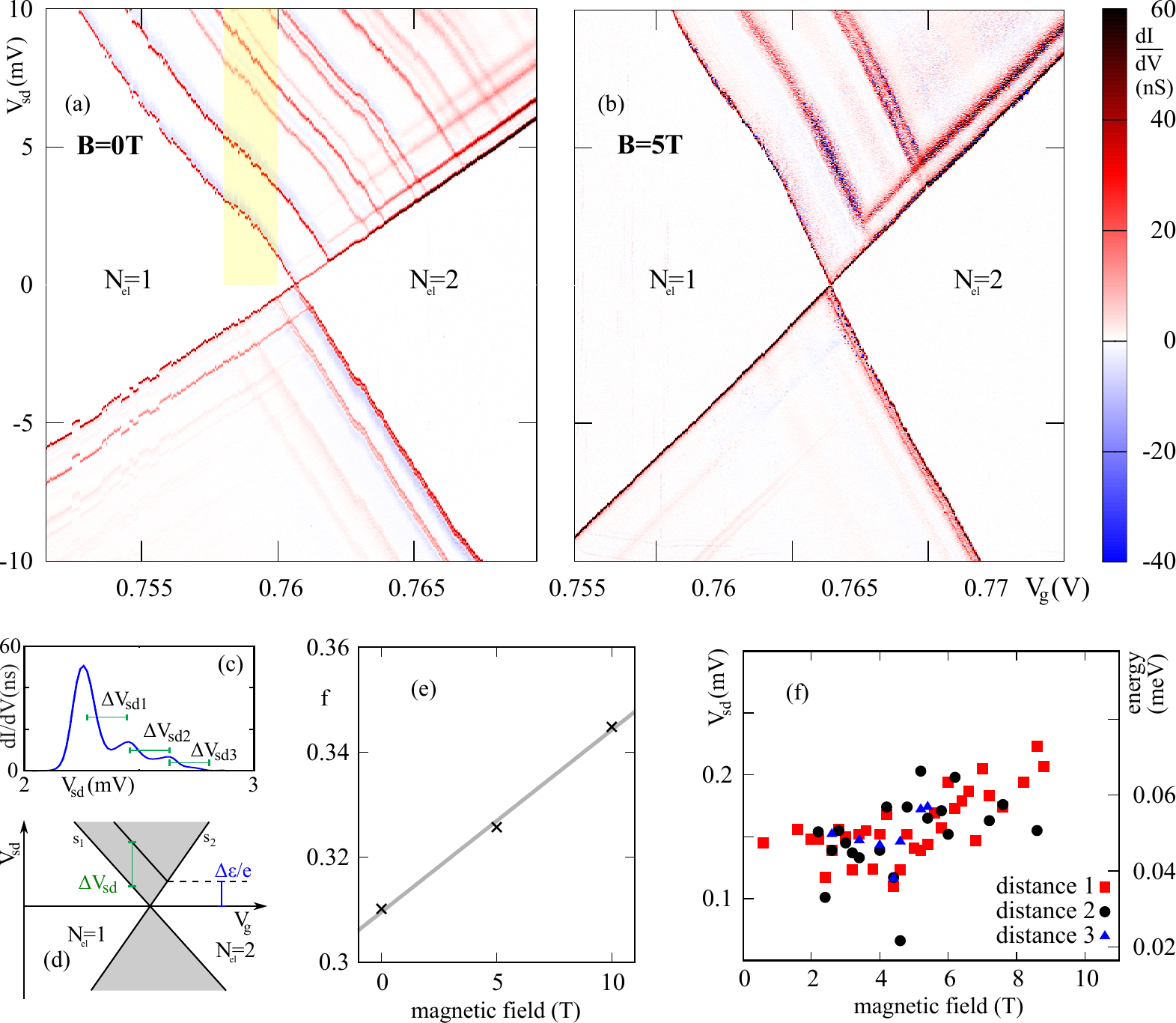}
\caption{
(a) Large bias voltage range stability diagram $\didv(\vg,\vsd)$ of the $1 
\le \nel \le 2$ transition at $\bpar=0\un{T}$. Fig.~\ref{fig:overview}(d) of 
the main text is a detail zoom of this plot. The parameter region of the trace 
cuts evaluated in Figs.~\ref{fig:groundstate} and \ref{fig:excitations} is 
shaded in yellow.
(b) Large bias voltage range stability diagram $\didv(\vg,\vsd)$ at 
$\bpar=5\un{T}$.
(c) Illustrative example trace $\didv(\vsd)$ for constant \bpar\ and \vg. 
(d) Using the slopes $\Delta \vsd / \Delta \vg$ of the two edges of the single 
electron tunneling region in the stability diagram, $s_1 < 0$ and $s_2 > 0$, the 
bias differences $\Delta \vsd$ can be converted to energy differences $\Delta 
\epsilon$ (here demonstrated for a $\nel=2$ state). 
(e) Conversion factor $f(\bpar)$, defined via $\Delta \epsilon = f(\bpar) \cdot 
e \Delta\vsd$, as extracted from the stability diagrams at $\bpar=0\un{T}$ 
(see (a), Fig.~\ref{fig:overview}(d)), $\bpar=5\un{T}$ (see (b)), and 
$\bpar=10\un{T}$ (see Fig.~\ref{fig:overview}(e)).
(f) Manually extracted distances between conductance peak positions, cf. (a), 
using the data set of Fig.~\ref{fig:groundstate}(b). Left axis: raw distances 
in bias voltage, right axis: distances converted to energy, using $f(5\un{T})$.
\label{fig:harmonic}}
\end{figure*}
Figures~\ref{fig:harmonic}(a) and (b) display the conductance near the $1 
\le \nel \le 2$ degeneracy point, for $\bpar=0\un{T}$ and $\bpar=5\un{T}$ and 
over a large bias voltage range $|\vsd|\le 10\un{mV}$. The magnetic field 
dependent measurements evaluated in Figs.~\ref{fig:groundstate} and 
\ref{fig:excitations} correspond to traces taken across the single electron 
tunneling region, in the parameter range indicated in 
Fig.~\ref{fig:harmonic}(a) by a yellow marker. Part of such a trace 
$\didv(\vsd)$ for constant \bpar\ and \vg\ is shown schematically in 
Fig.~\ref{fig:harmonic}(c). From it, the peak distances ${\Delta \vsd}_1$, 
${\Delta \vsd}_2$, ${\Delta \vsd}_3$ as indicated in the drawing can be 
extracted. They correspond to excitation energies $\Delta \epsilon_1$, $\Delta 
\epsilon_2$, $\Delta \epsilon_3$. 

In the stability diagrams, conductance lines ending at the $\nel=2$ Coulomb 
blockade region (i.e., for $\vsd>0$ with negative slope) correspond to 
2-electron excitations; lines ending at the $\nel=1$ Coulomb blockade region 
(i.e., for $\vsd>0$ with negative slope) correspond to 1-electron excitations. 
In the evaluated parameter region, only lines with negative slope are visible. 
For them, the conversion from bias voltage distances $\Delta \vsd$ to 
excitation energies $\Delta \epsilon$, based on the capacitances in the quantum 
dot system, can be illustrated by the sketch of Fig.~\ref{fig:harmonic}(b). 
Given the slopes $\Delta \vsd / \Delta \vg$ of the two edges of the single 
electron tunneling region in the stability diagram, $s_1 < 0$ and $s_2 > 0$, as 
indicated in the sketch, the conversion factor $f(\bpar, \vg)$ can be derived 
from elementary geometry as
\begin{equation}
 \Delta \epsilon = \frac{
 1
 }{
 1 -({s_1}/{s_2})
 }
 \; e\, \Delta \vsd = f(\bpar, \vg) \; e\, \Delta \vsd.
\end{equation}
Here we indicate with $f(\bpar, \vg)$ that the factor can change with magnetic 
field and gate voltage due to modification of the electronic wave function 
shapes and thus the charge distributions and capacitances. The value
of $f$, as extracted from stability diagrams at $\bpar = 0, 5, 10\un{T}$, is
plotted in Fig.~\ref{fig:harmonic}(c), with a linear fit added. In the plotted 
range, $f$ varies by only approximately $10\%$.

The result of a manual evaluation of the data of Fig.~\ref{fig:groundstate}(b), 
where peak positions have been read out from line plots, is shown in 
Fig.~\ref{fig:harmonic}(d). Here, the red squares correspond to the peak 
distance of the first sideband relative to the base condutance resonance, the 
black dots to the one of the second sideband relative to the first sideband, 
and the blue triangles to the one of the third sideband relative to the second 
sideband. The left axis displays distance in bias voltage, the right axis the 
value converted into energy, using constant $f(5\un{T})$ for simplicity. 
Within the scatter, the three types of points lie in the same band of values, 
indicating equidistant quantum states and thus harmonic oscillator behaviour.
The oscillator quantum is for $0\un{T} \le \bpar \le 6\un{T}$ approximately 
constant at $\Delta \vsd \simeq 0.15\un{mV}$ corresponding $\Delta \epsilon = 
\hbar \omega \simeq 50\,\mu\text{eV}$; this is the region predominantly 
discussed in the main text.  

For larger \bpar, the data points seem to indicate a gradual increase in 
$\Delta \epsilon$. Given the decreasing signal to noise ratio and limited data 
for this field range it is unclear whether this is a real effect. Surprisingly, 
the stability diagram at $\bpar=10 \un{T}$, Fig.~\ref{fig:overview}(e), 
displays $\Delta \vsd \simeq 0.25\un{mV}$, which would confirm such an increase.

The theoretical value for the energy quantum of the carbon nanotube
longitudinal vibration is given by \cite{prl-sapmaz-2006}
\begin{equation}
\Delta \epsilon_\text{th} = \frac{h}{L}\sqrt{\frac{Y}{\rho}},
\end{equation}
where $L$ is the nanotube length, $Y$ is Young's modulus, and $\rho$ is the
nanotube mass density. Assuming $\rho=1.3\un{g/cm$^3$}$ and $Y=1\un{TPa}$, this
results in \cite{prl-sapmaz-2006}
\begin{equation}
\Delta \epsilon_\text{th} \approx \frac{0.11\un{meV}}{L(\mu\text{m})}
\end{equation}
\begin{figure*}[t]
\centering\includegraphics[width=0.7\textwidth]{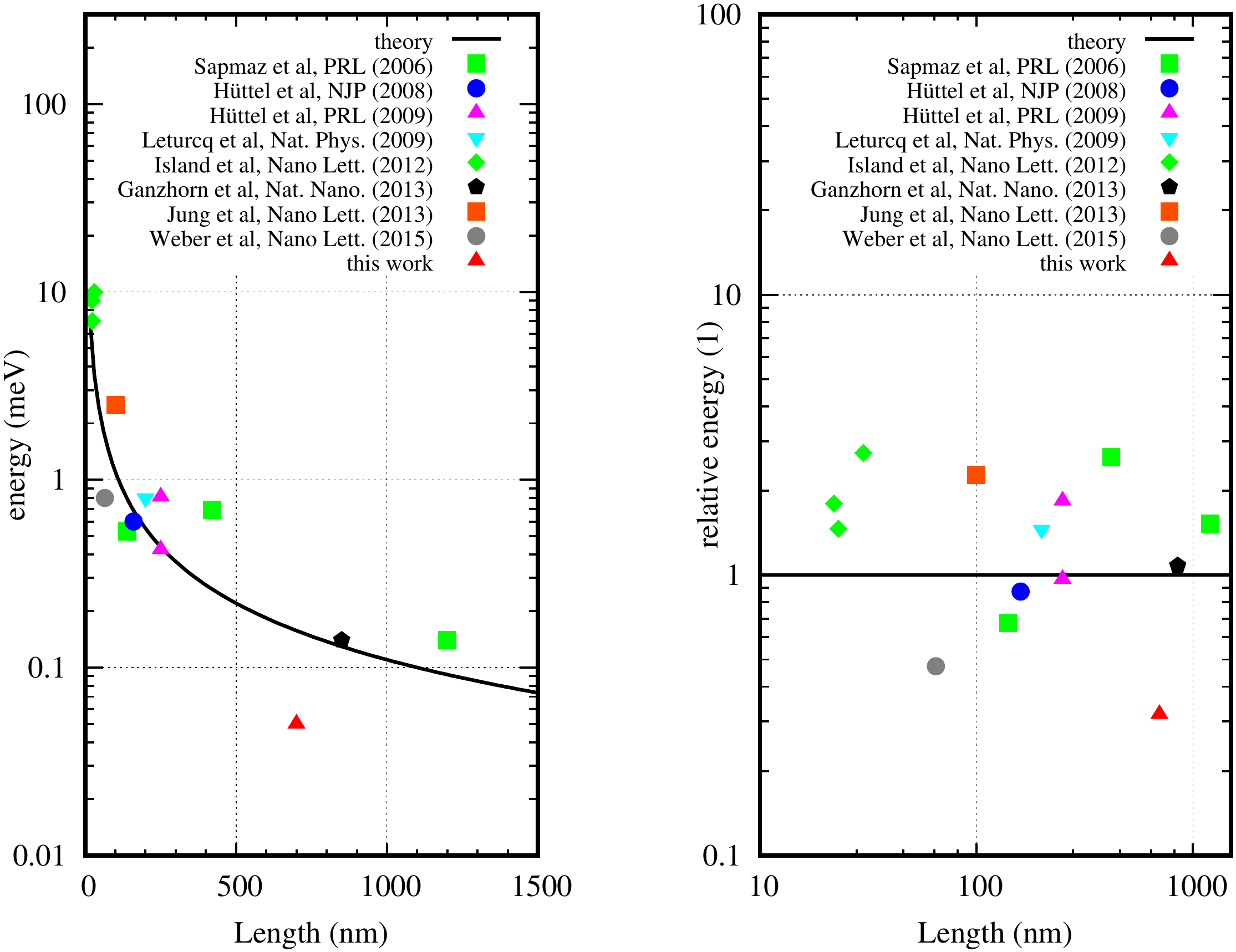}
\caption{Overview of carbon nanotube longitudinal vibration oscillator quanta
$\Delta \epsilon = \hbar\omega$ observed in published literature
\cite{prl-sapmaz-2006, quantumnems, cocoset, nphys-leturcq-2009, 
nl-island-2012, nnano-ganzhorn-2013, nl-jung-2013, nl-weber-2015}, as function 
of device length, in comparison with the theoretical result $\Delta 
\epsilon_\text{th}(L)= 110\un{meV}/L[\text{nm}]$. Left: unscaled $\Delta 
\epsilon(L)$, linear length scale; right: $\Delta \epsilon(L) / \Delta 
\epsilon_\text{th}(L)$, logarithmic length scale.
\label{fig:scattering}}
\end{figure*}
For our device, using the contact separation of $L=0.7\un{$\mu$m}$ as 
approximate value of the suspended nanotube segment length, we obtain $\Delta 
\epsilon_\text{th} \simeq 160\,\mu\text{eV}$. While there is a clear deviation, 
our measurement still lies at the edge of the typical scatter of oscillator 
quanta observed in experimental literature, see Fig.~\ref{fig:scattering} for 
an overview.

\section{Franck-Condon model and fit procedure}\label{a-fit}

In the context of single electron tunneling through a carbon nanotube, 
the Franck-Condon coupling parameter $g$ describes the spatial shift of the 
nanotube equilibrium position as harmonic oscillator when an additional 
electron is added to it. It is given by $g = ( {\Delta x_0}/{x_\text{zpf}} )^2 
/2$, where $\Delta x_0$ is the shift in equilibrium position, $\Delta x_0 = 
x_0(N+1) - x_0(N)$. In the denominator, $ x_\text{zpf} = \sqrt{{\hbar}/({m 
\omega})}$ is the characteristic length scale of the harmonic oscillator, 
describing the wave function extension of the ground state and/or its zero 
point fluctuations. For $\hbar\omega = 50\,\mu\text{eV}$ and $m=1.3\times 
10^{-21}\un{kg}$ \cite{magdamping}, we obtain $x_\text{zpf} = 1.0\un{pm}$.

At finite temperature, a harmonic oscillator can both absorb and emit vibrons.
Here, in the limit of low temperature and fast vibrational relaxation compared
to the tunnel rates, we assume that for any single electron tunneling process
we start out in the \nel\ electron vibrational ground state. For each number 
of vibration quanta $n$, there is a distinct overlap $\left| 
\braket{\Psi_0(\nel)\,}{\Psi_n(\nel+1)} \right|^2$ of the \nel\ electron 
vibrational ground state with an $\nel+1$ electron, $n$ vibron state. This 
leads to a series of equidistant steps in current $I(\vsd)$ or peaks in 
differential conductance $\didv(\vsd)$, whenever sufficient energy for reaching 
the next vibrational state becomes available.

The contributions of the vibrational states and thus the current step heights 
are given for the limit of low temperature and fast vibrational relaxation by 
the Poisson formula \cite{prb-braig-2003, prb-koch-2006},
\begin{equation}\label{eq:stepsize}
 \Delta I_n \propto \frac{e^{-g} g^n}{n!}, \quad n=0, 1, 2, \dots
\end{equation}
at an energy $\epsilon= n \hbar \omega$ from the bare electronic state 
transition.
\begin{figure}[tbp]
\centering
\includegraphics{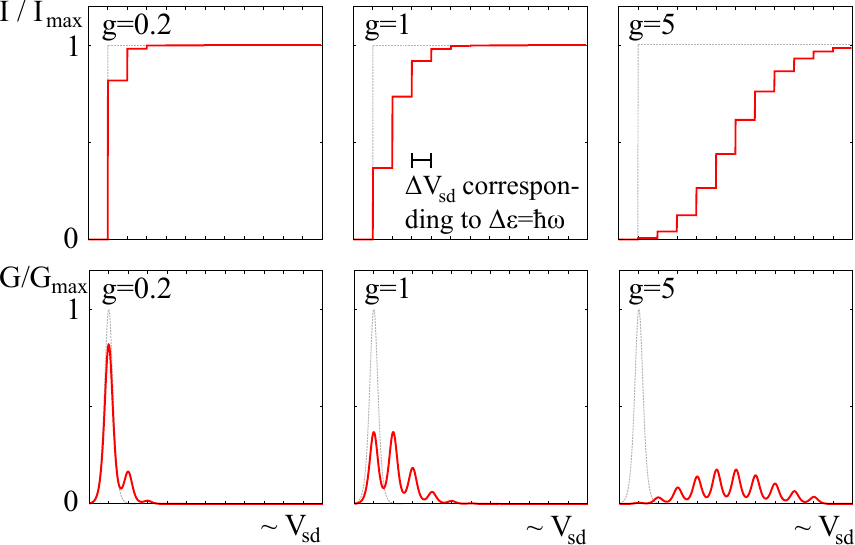}
\caption{
Sketch of the impact of the Franck-Condon coupling parameter $g$ on single 
electron tunneling: (a--c) current $I(\vsd)$, in an idealized system where no 
broadening (thermal, lifetime, or otherwise) is present, for (a) $g=0.2$, (b) 
$g=1$, and (c) $g=5$; (d--f) conductance at finite broadening, see the text. In 
each panel the case of no electron-vibron interaction ($g=0$) is overlaid as 
gray dashed line.) 
\label{fig:couplingeval}}
\end{figure}
The resulting step function of the current, in the idealized case of total 
absence of effects such as thermal or lifetime broadening, is shown in 
Fig.~\ref{fig:couplingeval}(a-c) for different values of $g$.

In a measurement, the steps will be broadenend due to finite temperature, 
finite lifetime of the involved quantum states, and additional mechanisms such 
as, e.g., potential fluctuations. As approximation we assume that this 
broadening equally affects all steps. Then, the differential conductance $G = 
\didv$ exhibits a sequence of peaks, each corresponding to one step in current, 
with conductance peak heights proportional to the current step heights. Even 
though $G$ is in our measurement a derived quantity, it is both a better base 
for visualization of the phenomena, see the figures of this work, and a better 
base for numerical curve fitting than the measured current. To minimize the 
impact of postprocessing, we use the for fits the raw numerically 
differentiated conductance $\Delta I/ \Delta \vsd$, without any smoothing or 
other numerical filtering applied. 

As already mentioned in the main text, we find that our data is fitted well
using the typical shape of a thermally broadened Coulomb oscillation, 
\begin{equation}
G_\text{th} (\vsd, V_\text{sd}^0)= 
\cosh^{-2}\left(\frac{\vsd-V_\text{sd}^0}{\gamma}\right),
\end{equation}
where $\gamma$ describes the peak width. For our fit model, we sum up a 
sequence of these peaks, weighted according to Eq.~\ref{eq:stepsize}, cutting 
off at $n=10$ since subsequent terms are negligible for $g\lesssim 1$, and 
normalize such that the maximum conductance $G_\text{max}$ at the electronic 
base resonance ($n=0$) can be used as convenient fitting parameter. This 
results in
\begin{equation}
 G(\vsd, V_\text{sd}^0) = G_\text{max}
\frac{
\sum_{n=0}^9 \; P_n G_\text{th}(\vsd, V_\text{sd}^0 + n\,\Delta\vsd)
}{
\sum_{n=0}^9 \; P_n G_\text{th}(V_\text{sd}^0, V_\text{sd}^0 + n\,\Delta\vsd)
}.
\end{equation}

Given the combination of large scatter in the data, see, e.g., 
Fig.~\ref{fig:groundstate}(c), and the large number of free fit parameters 
$\left( V_\text{sd}^0, G_\text{max}, \gamma, g, \Delta\vsd=\hbar\omega/(e f) 
\right)$ the fit displays instabilities and large error bars. However, it turns 
out that the fit results for both $\gamma$ and $\Delta\vsd$ scatter in the 
evaluated regions around field-independent values (also validating our manual 
evaluation of Fig.~\ref{fig:harmonic}(d)). As a consequence, we fix both 
parameters to an average value and re-run the fit with a reduced set of free 
parameters $\left( V_\text{sd}^0, G_\text{max}, g\right)$.
\begin{figure*}[p]
\centering
\includegraphics[width=0.8\textwidth]{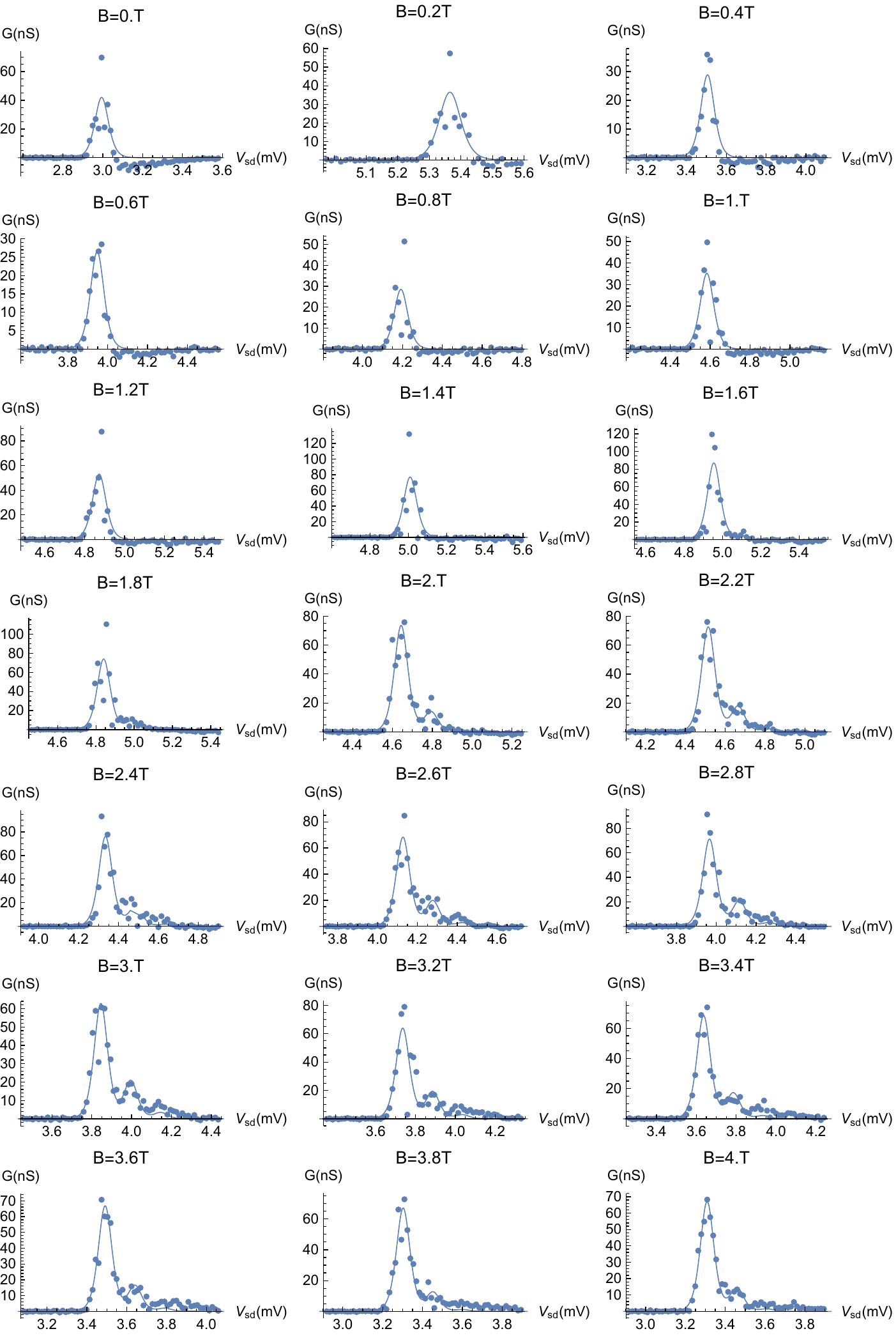}
\caption{
\label{fig:rawfits} Example plots of numerically obtained differential 
conductance corresponding to data traces of Fig.~\ref{fig:groundstate}(b), and 
the fits resulting in the parameters plotted in Fig.~\ref{fig:groundstate}(f) 
(blue squares there).}
\end{figure*}
To give an insight into the results of this procedure, Figure~\ref{fig:rawfits} 
shows for part of the evaluation both the raw data and the best fit functions; 
these correspond to data traces of Fig.~\ref{fig:groundstate}(b) and the fit 
prarameters plotted in Fig.~\ref{fig:groundstate}(f) as blue squares.

\section{Classification of excited state resonances}\label{app-classif}

In a magnetic field at an angle $\varphi$ to the carbon nanotube axis, the 
linearized single particle Hamiltonian of an electron in a carbon nanotube 
quantum dot close to the Dirac point is given by \cite{brokensu4, highfield}
\begin{multline}\label{eq:hamilton}
 \hat{H}_\text{CNT}\; = \;
\underbrace{\varepsilon_d\,\hat{I}_{\sigma}\otimes\hat{I}_{\tau}}_{
\text{``shell''}}
 \;
+ \; 
\underbrace{\frac{\DKK}{2}\hat{I}_{\sigma}\otimes\hat{\tau}_x}_{\text{valley 
mixing}}
 \;
+ \; 
\underbrace{\frac{\DSO}{2} \hat{\sigma}_z 
\otimes\hat{\tau}_z}_{\text{spin-orbit interaction}}
\;
+ \; \\
\;
+ \;
\underbrace{\frac{g_\text{s}\mu_\text{B}|\vec{B}|}{2}
 \left(\cos\varphi\,\hat{\sigma}_z + 
 \sin\varphi\,\hat{\sigma}_x\right) \otimes\hat{I}_{\tau}}_{\text{Zeeman 
effect}}
\;
+ \; \\
\;
+ \;
 \underbrace{g_\text{orb}\mu_\text{B}|\vec{B}| 
\cos\varphi\; 
 \hat{I}_{\sigma}\otimes\hat{\tau}_z}_{\text{orbital angular moment}}
\end{multline}
Here, the Pauli matrices $\hat{\sigma}_i$ act on the spin part of the wave 
function, and the Pauli matrices $\hat{\tau}_j$ on the orbital/valley part of 
the wave function, respectively; the basis is given by spin and valley 
eigenstates in axial direction.

For a field alignment parallel to the nanotube ($\varphi=0$), already at 
moderate field (e.g., $\bpar \sim 0.5\un{T}$ \cite{highfield}) the contribution 
of the orbital angular moment exceeds the valley mixing, and the eigenstates of 
the Hamiltonian become spin and valley eigenstates, with spin and orbital 
angular moment aligned parallel to the nanotube axis. With $\sigma=\pm 1$ and 
$\tau=\pm 1$ describing spin and orbital angular moment direction, the magnetic 
field dependent terms of Eq.~\ref{eq:hamilton} then lead to the energy 
contribution
\begin{equation}
 E_{\sigma\tau}(\bpar)= \left(
 \sigma \frac{g_\text{s}\mu_\text{B}}{2}
+ 
 \tau g_\text{orb}\mu_\text{B} \right) \bpar. 
\end{equation}
The dominant term is here the orbital contribution \cite{nature-minot-2004, 
rmp-laird-2015, highfield}. This can be clearly observed, e.g., in 
Fig.~\ref{fig:oneelectron}, where at large field the two visible conductance 
resonances correspond to two single-particle states with equal orbital angular 
moment but opposite spin. As a consequence, we go one step further and classify 
our states only by orbital angular momentum direction (corresponding to K' or K 
valley), i.e. by dominant magnetic field dispersion
\begin{equation}
 E_{\tau}(\bpar)= \tau g_\text{orb}\mu_\text{B} \bpar, \quad \tau=\pm 
1
\end{equation}

In the case of the second single electron tunneling region where the charge 
occupation is $1\le \nel \le 2$, resonance lines correspond to energy 
differences $\Delta E_{12}(\bpar) = E_2(\bpar) - E_1(\bpar)$, and the 
dispersions in a magnetic field correspondingly to the orbital angular momentum 
{\em change} when a second electron tunnels onto the quantum dot, proportional 
to 
\begin{equation}
\frac{\text{d}(\Delta E_{12}(\bpar))}{\text{d}\bpar} = 
\frac{\text{d}(E_2(\bpar) - E_1(\bpar))}{\text{d}\bpar}.
\end{equation}
If the angular momentum of the first electron remains unchanged when the 
second eletron tunnels onto the dot, only two slopes corresponding to addition 
of a K' or K electron remain possible. From this we can classify the 
conductance lines of Fig.~\ref{fig:excitations}(a). In the low-bias region of 
the figure, two dominant slopes clearly emerge; we identify these with the 
addition of a K'-valley electron (downward slope) or a K-valley electron 
(upward slope), respectively \cite{nature-minot-2004, rmp-laird-2015,
highfield}.

\section{Evaluating the sidebands of excited states}
\label{app-excited}

The data points of Fig.~\ref{fig:updown} are obtained as follows. We separate 
the conductance resonance pattern into line segments, each terminated by line 
crossings or anticrossings. On each segment that conforms to one of the two 
dominant slopes, we select a point $(\bpar, \vsd)$ near the high-\bpar\ end of 
the segment; these points are the locations of the symbols in 
Fig.~\ref{fig:updown}(a). The five traces $G(\vsd)$ closest to this point in 
magnetic field are evaluated using the fit procedure as described above. To 
reduce the impact of outliers, the plotted values in Fig.~\ref{fig:updown} are 
then given by the median of the fit parameters $g$ from these five evaluations.

\section{Variable coupling model}
\label{app-toy}

One of the key results of \cite{highfield}, based on \cite{prb-akhmerov-2008, 
rmp-castroneto-2009, prb-marganska-2011}, is that a magnetic field parallel to 
the axis of a carbon nanotube does not only modify the circumferential 
electronic wave function via an Aharonov-Bohm phase, but also the wave function 
in axial direction. While a typical one-dimensional ``quantum box'', i.e., a 
potential well with hard walls of infinite height, forces the wave function of a 
trapped particle to zero at both ends, this construction is not possible for 
carbon nanotubes. With the exception of armchair nanotubes, there are no 
solutions that allow the electronic wave function to become zero on both ends 
for both graphene sublattices. Instead, the longitudinal quantization can be  
given by the condition that the wave function on each sublattice becomes zero 
at the end where that particular sublattice provides the majority of boundary 
atoms. With $L$ as the length of the nanotube segment, this introduces a 
so-called cross-quantization condition \cite{highfield}, 
\begin{equation}
\label{eq:cross-quantization}
e^{2i\kappa_\parallel L} \overset{!}{=}
 \frac{\tau \kappa_\perp + i\kappa_\parallel}{\tau \kappa_\perp - 
i\kappa_\parallel}.
\end{equation}
The axial wave number $\kappa_\parallel$ and the circumferential wave number
$\kappa_\perp$ of an electronic state are coupled; one example solution is 
sketched in Fig.~\ref{fig:toymodel}(a) following \cite{highfield}. The axial 
magnetic field tunes the longitudinal profile and charge distribution of the 
electronic wave function from a $\lambda/4$ resonator like shape, with a finite 
value at one edge, near $\bpar=0$ all the way to a $\lambda/2$ resonator like 
shape at high field; see the schematic drawings in Fig.~\ref{fig:toymodel}(a). 

In \cite{highfield}, wave function envelopes have been calculated numerically 
using a tight-binding approach. Here, we approximate these results using a 
simple functional dependence: for a quantum dot of length 1, we 
write for sublattice A
\begin{equation}
 \psi_A(x, \phi) = \sin(\pi \,\phi\, x), \quad x\in [0,1]
\end{equation}
and accordingly for sublattice B
\begin{equation}
 \psi_B(x, \phi) = \psi_A(1-x, \phi)
\end{equation}
The parameter $\phi$ describes the number of wave function antinodes along 
the quantum dot, i.e., in the range $x\in[0,1]$. Half-wavelength resonator 
solutions are given by $\phi=1,2,\dots$, quarter-wavelength resonator solutions 
are given by $\phi=0.5, 1.5, 2.5, \dots$. As can be seen in 
Fig.~\ref{fig:toymodel}(a), 
increasing a magnetic field from zero continuously tunes $\phi$, for a K' state 
from $\phi \approx 1.25$ towards $\phi=2$, for a K state from $\phi \approx 
1.25$ towards $\phi=1$.

\begin{figure*}
\centering
\includegraphics{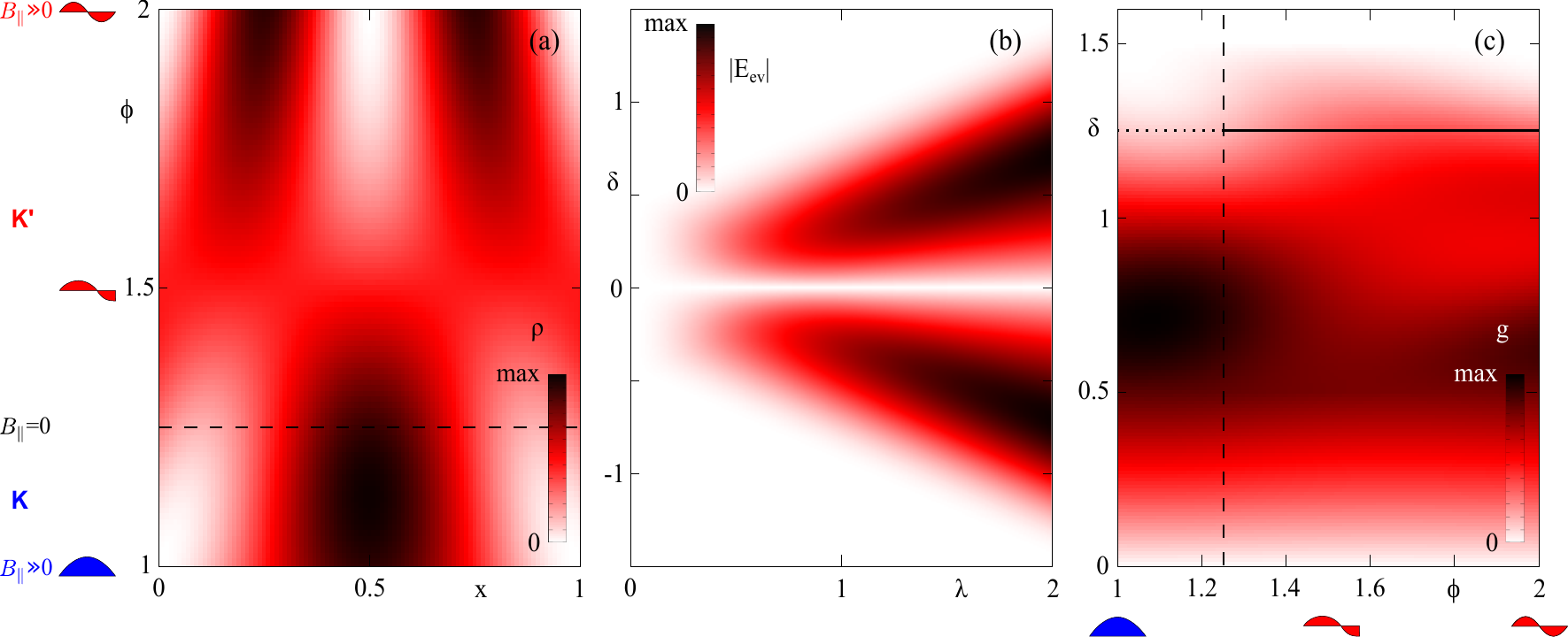}
\caption{Model construction: (a) Assumed linear single particle charge 
density $\rho$ along a nanotube quantum dot of length 1, as function of 
position $x$ and number of antinodes of the graphene sublattice wavefunction 
$\phi$, cf. \cite{highfield}. The approximate situation at low magnetic field 
is marked with a dashed line at $\phi=1.25$; $\phi=2$ corresponds to the high 
field limit of the K' state, $\phi=1$ to the high field limit of the 
corresponding K state. (b) Absolute value of the interaction energy $\left| 
E_\text{ev} \right|$ for $\phi=1$, as function of length ratio $\lambda = 
L_v/L_d$ and relative position of the centres $\delta = (x_v-x_d) / L_d$ of 
vibron and quantum dot. (c) Franck-Condon parameter $g$, approximated 
as $g\simeq E_\text{ev}^2$, for fixed $\lambda=2$ and as function of $\phi$ and 
$\delta$. The solid/dotted line at $\delta=1.25$ corresponds to the line cuts 
shown in Fig.~\ref{fig:toymodel}(c); $\phi=1.25$ is again marked with a dashed 
line.
\label{fig:modelstory}}
\end{figure*}
To obtain the total linear electron density, we now sum up the densities on the 
two sublattices, while normalizing such that a change in $\phi$ does not modify 
the total charge, 
\begin{equation}
\rho(x, \phi)=\frac{
\left|\psi_A(x, \phi)\right|^2 + \left|\psi_B(x, \phi)\right|^2
}{
\int\limits_0^1 
\left|\psi_A(x, \phi)\right|^2 + \left|\psi_B(x, \phi)\right|^2\,\text{d}x
}
\end{equation}
The result is plotted in Fig.~\ref{fig:modelstory}(a). 

Reference~\cite{njp-donarini-2012} discusses the electron-vibron interaction of 
an interacting electron system using a Tomonaga–Luttinger model. From the 
interaction Hamiltonian used there, Eq.~(12) in \cite{njp-donarini-2012}, 
coupling the electron density to the deformation potential, we take inspiration 
of our interaction energy, 
\begin{equation}
 E_\text{ev}(\phi) \propto \int 
\rho(x,\phi))\frac{\text{d}u}{\text{d}x}\,\text{d}x
\end{equation}
where $u(x)$ is the vibration envelope of the mechanical mode. In addition, 
also following \cite{njp-donarini-2012} and its nomenclature, we allow for the 
vibron and the quantum dot to be shifted relative to each other and to be of 
different length; we parametrize this using length ratio $\lambda = L_v/L_d$ 
and relative position of the centres $\delta = (x_v-x_d )/L_d$ of vibron and 
quantum dot. Fig.~\ref{fig:modelstory}(b) plots for $\phi=1$ the interaction 
energy $\left|E_\text{ev}(\lambda,\delta)\right|$, reproducing basic features 
of the corresponding Tomonaga–Luttinger model plot, left panel of Fig.~5 in 
\cite{njp-donarini-2012}.

Given the comparatively small vibrational energy quantum in our device, cf. 
Fig.~\ref{fig:scattering}, and the strong electrostatic confinement of the 
quantum dot, we consider it likely that the quantum dot is smaller than the 
vibron envelope, i.e., $\lambda > 1$. We are now interested in the magnetic 
field induced modification of the Franck-Condon parameter $g \propto 
E_\text{ev}^2$ \cite{njp-donarini-2012}, and thus plot in 
Fig.~\ref{fig:modelstory}(c) this value as function of $\phi$ and $\delta$, for 
a fixed $\lambda=2$. The solid/dotted line corresponds to $\delta=1.25$, the 
value selected for the trace cut in Fig.~\ref{fig:toymodel}(c).

\end{document}